\documentclass[acmsmall, authorversion]{acmart}

\usepackage[usestackEOL]{stackengine}
\usepackage{xcolor}
\usepackage{multirow}

\usepackage{lscape}

\usepackage[english=american,threshold=40,thresholdtype=words]{csquotes}

\setcopyright{acmcopyright}
\copyrightyear{2024}
\acmYear{2024}
\acmDOI{}
\acmISBN{}

\acmConference[CSCW '24]{Conference on Computer-Supported Cooperative Work \& Social Computing}{November 09--13}{San José, Costa Rica}
\acmBooktitle{Conference on Computer-Supported Cooperative Work \& Social Computing (CSCW '24)}
\begin{document}

\strutlongstacks{T}


\title[RAI Guidelines]{RAI Guidelines: Method for Generating Responsible AI Guidelines Grounded in Regulations and Usable by (Non-)Technical Roles}

\author{Marios Constantinides}
\affiliation{%
  \institution{Nokia Bell Labs}
  \city{Cambridge}
  \country{United Kingdom}
}
\email{marios.constantinides@nokia-bell-labs.com}

\author{Edyta Bogucka}
\affiliation{%
  \institution{Nokia Bell Labs}
  \city{Cambridge}
  \country{United Kingdom}
}
\email{edyta.bogucka@nokia-bell-labs.com}

\author{Daniele Quercia}
\affiliation{%
  \institution{Nokia Bell Labs}
  \city{Cambridge}
  \country{United Kingdom}
}
\email{daniele.quercia@nokia-bell-labs.com}

\author{Susanna Kallio}
\affiliation{%
  \institution{Nokia}
  \city{Espoo}
  \country{Finland}
}
\email{susanna.kallio@nokia.com}

\author{Mohammad Tahaei}
\affiliation{%
  \institution{Nokia Bell Labs}
  \city{Cambridge}
  \country{United Kingdom}
}
\email{smh.tahaei@gmail.com}

\renewcommand{\shortauthors}{Trovato and Tobin, et al.}

\begin{abstract}
Many guidelines for responsible AI have been suggested to help AI practitioners in the development of ethical and responsible AI systems. However, these guidelines are often neither grounded in regulation nor usable by different roles, from developers to decision makers. To bridge this gap, we developed a four-step method to generate a list of responsible AI guidelines; these steps are: (1) manual coding of 17 papers on responsible AI; (2) compiling an initial catalog of responsible AI guidelines; (3) refining the catalog through interviews and expert panels; and (4) finalizing the catalog. To evaluate the resulting 22 guidelines, we incorporated them into an interactive tool and assessed them in a user study with 14 AI researchers, engineers, designers, and managers from a large technology company. Through interviews with these practitioners, we found that the guidelines were grounded in current regulations and usable across roles, encouraging self-reflection on ethical considerations at early stages of development. This significantly contributes to the concept of `Responsible AI by Design'— a design-first approach that embeds responsible AI values throughout the development lifecycle and across various business roles.
\end{abstract}




% ---> artificial intelligence, machine learning, are already included in the above keyweods, these are additional


% \textbf{Responsible AI by Design.}
\begin{teaserfigure}
  \includegraphics[width=\textwidth]{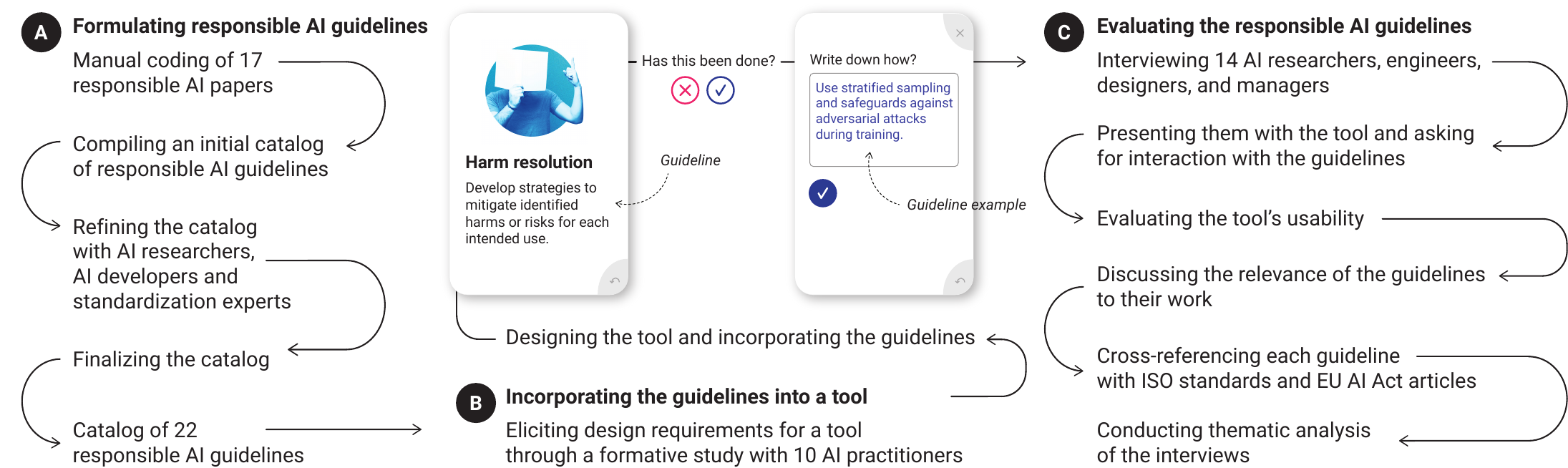}
  \caption{Overview of our method for generating responsible AI guidelines and evaluating them: (A) formulating responsible AI guidelines that are grounded in regulations and are usable by different roles; (B) incorporating the guidelines into a tool; and (C) evaluating them.}
  \label{fig:card-elements}
\end{teaserfigure}

\maketitle
\section{Introduction}
\label{sec:introduction}
The development of responsible AI systems~\cite{shneiderman2021responsible, kissinger2021age, tahaei2023human} has become a significant concern as AI technologies continue to permeate various aspects of society~\cite{shneiderman2022human}. While AI holds the potential to benefit humanity, concerns regarding biases~\cite{baeza2018bias, cramer2019translation, buolamwini2018gender} and the lack of transparency and accountability~\cite{mitchell2019model, rakova2021responsible} hinder its ability to unlock human capabilities on a large scale. In response, AI practitioners\footnote{We use the term practitioners to cover a wide range of stakeholders including AI engineers, developers, researchers, designers, ethics experts.} are actively exploring ways to enhance responsible AI development and deployment. One popular approach is the use of tools such as checklists~\cite{madaio2020co} or guideline cards~\cite{elsayed2023responsible, humanAIGuidelines2019, FeministDeck_2022} that are designed to promote AI fairness, transparency, and sustainability. These tools provide practical frameworks that enable practitioners to systematically assess and address ethical considerations throughout the AI development lifecycle. By incorporating checklists and guideline cards into their workflows, practitioners can evaluate key aspects such as data sources, model training, and decision-making processes to mitigate potential biases, ensure transparency, and promote the long-term sustainability of AI. However, these tools face two main challenges, creating a mismatch between their potential to support ethical AI development and their current design.

The first challenge is that these tools often exhibit a static nature, lacking the ability to dynamically incorporate the latest advancements in responsible AI literature and international standards~\cite{nist2023aiRisk, fjeld2020principled}. In the rapidly evolving field of responsible AI, new ethical considerations and regulatory guidelines constantly emerge (e.g., the EU AI Act~\cite{eu_ai_act_2022}). It is therefore crucial for AI practitioners to stay updated of these developments to ensure their AI systems align with the current ethical and responsible AI practices. While checklists and guideline cards are increasingly used to assist and enhance the development of responsible AI systems, they are rarely grounded in current regulations. For example, Vakkuri \emph{et al.}~\cite{eccolaCards2021} proposed the ECCOLA cards that are based on AI ethics guidelines (e.g., IEEE Ethically Aligned Design and EU Trustworthy AI), which are not meant to be grounded on any specific regulations. Additionally, guidelines can quickly become outdated (e.g., the AI Blindspots deck has undergone several iterations~\cite{AIBlindspotsV1, AIBlindspotsV2}), limiting their effectiveness in addressing evolving concerns related to fairness, transparency, and accountability.

The second challenge is that, while these tools emphasize the importance of involving stakeholders from diverse roles and backgrounds, they are often designed for specific AI practitioners (e.g., ML engineers), neglecting a broader spectrum of stakeholders (e.g., non-technical roles). Balayn \emph{et al.}~\cite{balayn2023fairness} found that less experienced practitioners in machine learning tend to use a limited set of metrics and methods from toolkits. Similarly, Deng \emph{et al.}~\cite{deng2022exploring} stressed the lack of standardized guidelines in toolkits like AIF360 for introducing fairness issues to non-technical collaborators. Therefore, it is important that toolkits enhance communication, provide comprehensive guidance and support for cross-functional collaboration~\cite{yildirim2023investigating}.

To overcome these challenges, we developed a four-step method to generate a list of responsible AI guidelines which we then incorporated in a tool to evaluate them (Figure~\ref{fig:card-elements}). With this method, we aim to equip different roles with actionable guidelines that are grounded in regulations. To achieve this, we focused on answering this main research question: \emph{How to generate responsible AI guidelines that are grounded in regulations and are usable by different roles?} In addressing this question, we made two main contributions\footnote{The project's site is at \url{https://social-dynamics.net/rai-guidelines}}: 

\begin{enumerate}
\item We proposed a four-step method for generating Responsible AI guidelines; these steps are: \emph{(1)} manual coding of 17 papers on responsible AI; \emph{(2)} compiling an initial catalog of responsible AI guidelines; \emph{(3)} refining the catalog through interviews with 10 AI researchers and engineers, and workshops with 4 standardization experts; and \emph{(4)} finalizing the catalog. This procedure resulted into a set of 22 Responsible AI guidelines (\S\ref{sec:method}).

\item We evaluated the 22 guidelines in a user study with 14 AI researchers, engineers, designers, product managers from a large technology company (\S\ref{sec:userstudy}) by designing and deploying a tool incorporating the guidelines. To develop the tool, we conducted a formative study with 10 AI practitioners to determine key design requirements. Using these requirements, we populated the tool with the guidelines and conducted the case study. Interviews with the 14 AI researchers, engineers, designers, and managers revealed that the guidelines were grounded in current regulations and were effectively usable across different roles, promoting self-reflection on ethical considerations in early development stages.
\end{enumerate}

In light of these findings, we discuss how our method contributes to the idea of ``Responsible AI by Design'' by contextualizing the guidelines, informing existing or new theories, and offering practical recommendations for incorporating responsible AI guidelines into toolkits, and
recommendations for technical and non-technical roles in enabling organizational accountability (\S\ref{sec:discussion}).

\section{Related Work}
\label{sec:related}

We surveyed various lines of research that our work draws upon, and grouped them into two main areas: \emph{(1)} AI regulation and governance (\S\ref{subsec:ai_governance}), and \emph{(2)} responsible AI practices and toolkits (\S\ref{sec:sub-raipractices}). 

\subsection{AI Regulation and Governance}
\label{subsec:ai_governance}
The landscape of AI regulation and governance is constantly evolving~\cite{jobin2019global, mittelstadt2016ethics}. At the time of writing, the European Union (EU) has endorsed new transparency and risk-management rules for AI systems known as the EU AI Act~\cite{eu_ai_act_2022}, which is expected to become law in 2024. Similarly, the United States (US) has recently passed a blueprint of the AI Bill of Rights in late 2022~\cite{us_ai_bill}. This bill comprises \emph{``five principles and associated practices to help guide the design, use, and deployment of automated systems to protect the rights of the American public in the age of AI.''} Both the EU and US share a conceptual alignment on key principles of responsible AI, such as fairness and explainability, as well as the importance of international standards (e.g., ISO 24028 for Trustworthiness). 

%the specific AI risk management regimes they are developing are potentially diverging, creating an ``artificial divide''~\cite{ecfr}. The EU aims to become the leading regulator for AI globally, while the US takes the view that excessive regulation may impede innovation.

Notable predecessors to AI regulations include the EU GDPR law on data protection and privacy~\cite{eu_gdpr}, the US Anti-discrimination Act~\cite{us_anti_discrimination}, and the UK Equality Act 2010~\cite{uk_equality}. GDPR's Article 25 mandates that data controllers must implement appropriate technical and organizational measures during the design and implementation stages of data processing to safeguard the rights of data subjects. The Anti-discrimination Act prohibits employment decisions based on an individual's race, color, religion, sex (including gender identity, sexual orientation, and pregnancy), national origin, age (40 or older), disability, or genetic information. This legislation ensures fairness in AI-assisted hiring systems. Similarly, the UK Equality Act provides legal protection against discrimination in the workplace and wider society.

The National Institute of Standards and Technology (NIST), a renowned organization for developing frameworks and standards, recently published an AI risk management framework~\cite{nist2023aiRisk}. According to the NIST framework, an AI system is defined as \emph{``an engineered or machine-based system capable of generating outputs such as predictions, recommendations, or decisions that influence real or virtual environments, based on a given set of objectives. These systems are designed to operate with varying levels of autonomy.''} Similarly, the Principled Artificial Intelligence white paper from the Berkman Klein Center~\cite{fjeld2020principled} highlights eight key thematic trends that represent a growing consensus on responsible AI. These themes include privacy, accountability, safety and security, transparency and explainability, fairness and non-discrimination, human control of technology, professional responsibility, and the promotion of human values. Building on these themes, previous works have proposed a set of guidelines involving specific groups of AI practitioners. Saleema \emph{et al.}~\cite{humanAIGuidelines2019} proposed 168 guidelines on how to design AI tailored to HCI practitioners. Similarly, Vakkuri \emph{et al.}~\cite{eccolaCards2021} formulated AI ethics guidelines tailored to researchers and technologists. No subsequent work has associated these guidelines with current international standards or regulations.
\smallskip

\noindent\textbf{Research Gaps.} As AI regulation and governance continue to evolve, AI practitioners are faced with the challenge of staying updated not only with the changing guidelines, but also with regulations, requiring significant time and effort. Because prior guidelines lacked alignment with regulations, standards, and the input of experts in those fields, this work aims to create a methodology for crafting responsible AI guidelines that adhere to regulations and standards.

\subsection{Responsible AI Practices and Toolkits}
\label{sec:sub-raipractices}

\noindent\textbf{Responsible AI Toolkits.} At the time of writing, the OECD's website lists 613 toolkits dedicated to fostering the development and deployment of responsible AI systems~\cite{oecd_rai_toolkits}. These toolkits are essential for operationalizing guidelines and regulations to assist AI practitioners such as engineers and researchers in addressing algorithmic bias~\cite{bird2020fairlearn, gebru2021datasheets}, explaining algorithmic decisions~\cite{arya2019one}, and ensuring privacy in AI systems~\cite{fjeld2020principled}. For addressing algorithmic bias, Google's Fairness Indicators toolkit allows developers to assess data distribution and model performance across user-defined groups~\cite{google2022fairness}. IBM's AI Fairness 360 offers fairness metrics for bias mitigation~\cite{ibm2022ai}. Microsoft's Fairlearn assesses model impact on specific groups (e.g., under-represented populations) in terms of fairness and accuracy~\cite{fairlearn2022}. For explaining algorithmic decisions, IBM's AI Explainability 360 provides metrics and guidance for explainability, and new visualization techniques to enhance transparency~\cite{ibm2019_XAI, google2022pair, nvidia2022}. Finally, for ensuring privacy in AI systems, IBM's AI Privacy 360 helps assess and mitigate privacy risks through data anonymization and minimization~\cite{cavoukian2009privacy, cavoukian2010privacy, fjeld2020principled}. 
\smallskip

\noindent\textbf{Toolkits Used in Practice.} Developing toolkits specialized for certain audiences such as AI developers can lead to techno-solutionism, focusing exclusively on technical fixes. However, responsible AI entails broader socio-technical challenges (e.g., diversity and inclusion in decision-making) that require involvement of different roles with diverse expertise and background~\cite{selbst2019}, and such an involvement is typically discussed in venues with a long-standing commitment to human-centered design such as CHI, CSCW, AIES, and FAccT.

Different roles (e.g., data scientists, ML engineers and developers, UX designers) use toolkits in various ways. Data scientists often struggle to fully grasp visualizations of interpretable tools (e.g., InterpretML~\cite{interpret_ml} and SHAP~\cite{lundberg2017unified}), hindering their ability to understand datasets and underlying models~\cite{kaur2020interpreting}. Experienced ML developers and engineers often go beyond what fairness toolkits offer to tackle algorithmic unfairness, while those with less experience typically use only a few metrics and methods from these toolkits~\cite{balayn2023fairness}. UX designers often rely on custom prototypes and their own past experiences to help contextualize responsible AI issues for non-technical colleagues~\cite{deng2022exploring} due to communication gaps~\cite{yildirim2023investigating}.

Major communication gaps between technical and non-technical roles typically arise because these roles are involved in different stages of a project, which is likely to create fragmentation in communication~\cite{piorkowski2021ai}. By exploring how data science teams collaborate, Zhang \emph{et al.}~\cite{zhang2020data} found that non-technical roles play more prominent roles in the early and late stages of projects, while technical roles primarily handle the core data and modeling tasks. However, this disparity in involvement at various project stages is likely to create fragmentation. In fact, Organizational Science research reinforces the notion that effective communication and collaboration is crucial for overcoming the ``silo mentality''~\cite{forbes_silo}. Due to this fragmentation and a lack of robust organizational support, practitioners often take on ``bridging'' roles to help the communication between the technical and non-technical project members~\cite{deng2023investigating}. One way of doing so is through ``leaky abstractions''~\cite{subramonyam2022solving}. These are representations that are meant to communicate the inner workings and technical aspects of an AI system to these roles. Similarly, Nahar \emph{et al.}~\cite{nahar2022collaboration} highlighted the extreme difficulty faced by non-technical practitioners in eliciting requirements due to the absence of suitable tools and the involvement of diverse stakeholders, highlighting the need for integrating communication features into toolkits. The design of such features was explored by Elsayed-Ali \emph{et al.}~\cite{elsayed2023responsible} who developed question cards to facilitate stakeholder group discussions. These cards included built-in mechanisms for the automatic and cyclical assignment of cards to different participants, ensuring that everyone had the opportunity to share their opinions during the discussion. \\ 

\noindent\textbf{Research Gaps.} While many toolkits emphasize the importance of involving stakeholders from diverse roles and backgrounds, they are frequently designed for specific stakeholders (e.g., ML engineers), thereby neglecting a broader spectrum of roles (e.g., non-technical). To address this gap, we aim to develop a set of actionable guidelines that are usable by a diverse range of stakeholders.
\section{Author Positionality Statement}
\label{sec:positionality}
Understanding researcher positionality is crucial for transparently examining our perspectives on methodology, data collection, and analyses~\cite{frluckaj2022gender, havens2020situated}. In this paper, we situate ourselves in a Western country during the 21\textsuperscript{st} century, writing as authors primarily engaged in academic and industry research. Our team comprises three males and two females from Southern, Eastern, and North Europe, and Middle East with diverse ethnic and religious backgrounds. Our collective expertise spans various fields, including human-computer interaction (HCI), ubiquitous computing, software engineering, artificial intelligence, data visualization, and digital humanities.

It is important to recognize that our backgrounds and experiences have shaped our positionality. As HCI researchers affiliated with a Western organization, we acknowledge the need to expand the understanding of the research questions and methodology presented in this paper. Consequently, our positionality may have influenced the subjectivity inherent in framing our research questions, selecting our methodology, designing our study, and interpreting and analyzing our data.

% It is important to recognize that our backgrounds and experiences have shaped our positionality. As HCI researchers affiliated with a predominantly Western organization, we acknowledge the need to expand the understanding of the research questions and methodology presented in this paper. Consequently, our positionality may have influenced the subjectivity inherent in framing our research questions, selecting our methodology, designing our study, and interpreting and analyzing our data.

\section{Method for Generating Responsible AI Guidelines}
\label{sec:method}
\begin{figure}[t!]
  \centering
  \includegraphics[width=1.0\textwidth]{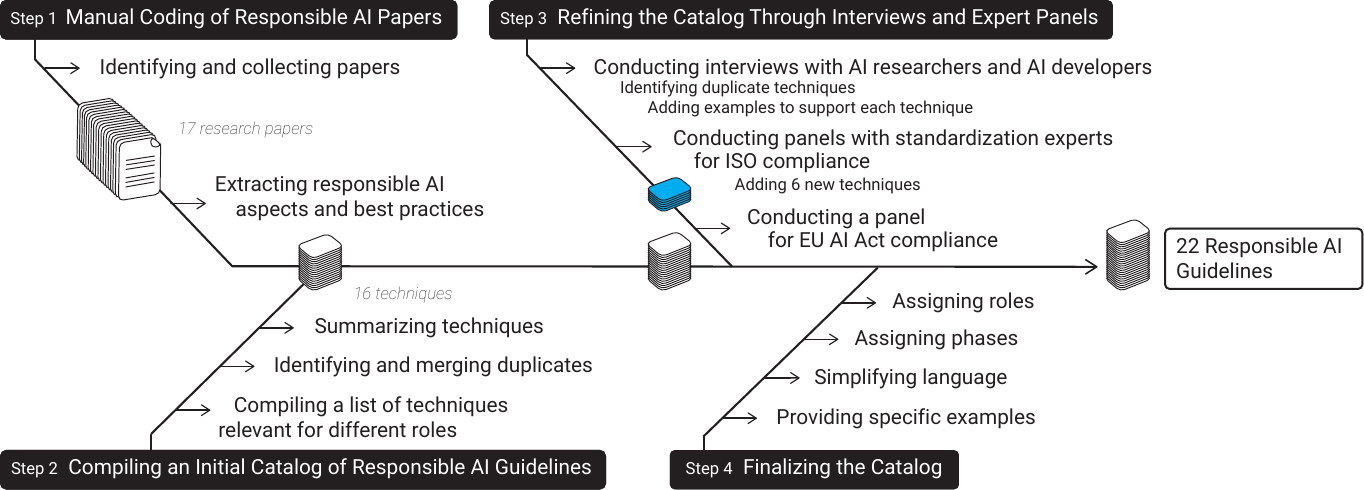}
  \caption{Four-step method for generating responsible AI guidelines. These guidelines were derived from research papers, and are in line with ISO standards and the EU AI Act~\cite{eu_ai_act_2022}.}
  \label{fig:steps}
\end{figure}

To generate a list of responsible AI guidelines, we followed a four-step process (Figure~\ref{fig:steps}), based on the methodology proposed by Michie \emph{et al.}~\cite{michie2013behavior}. This process allowed us to identify the essential element of a guideline, referred to as the ``active ingredient,'' focusing on the ``what'' rather than the ``how''~\cite{michie2011strengthening}. A similar parallel can be drawn in software engineering, where the ``what'' represents the software requirements and the ``how'' represents the software design, both of which are important for a successful software product~\cite{aggarwal2005software}. However, by shifting the focus to the ``what,'' AI practitioners can develop a clearer understanding of the objectives and goals they need to achieve, fostering a deeper comprehension of complex underlying ethical concepts. Throughout this process, we actively engaged a diverse group of stakeholders, including AI engineers, researchers, designers, product managers, and experts in law and standardization. As a result, we were able to formulate a total of \textbf{22} responsible AI guidelines (Panel A of Figure~\ref{fig:card-elements}). \\

\subsection{Manual Coding of Responsible AI papers}
\label{subsec:step1}
In the first step, we compiled a list of key scientific articles focusing on responsible AI guidelines applicable to a diverse set of roles, and manually coded them. We created this list by targeting papers published in renowned computer science conferences, such as the ACM CHI, CSCW, FAccT, AAAI/ACM Conference on AI, Ethics, and Society (AIES), and scientific literature from the medical domain (e.g., the Annals of Internal Medicine). Note that we did not conduct a systematic literature. Instead, we identified 17 papers that served as a starting point to compile an initial catalogue of techniques covering a broad range of responsible AI aspects, including fairness, explainability, sustainability, and best practices for data and model documentation and evaluation. These are foundational papers in responsible AI, and, as we shall see in a subsequent step of our methodology (\S\ref{sec:step_adding_iso}), we refined the techniques identified from those papers through interviews and expert panels as well as cross-referencing them with the EU AI Act and ISO standards.

These papers encompass a growing body of research focusing on the work practices (e.g., ensuring fairness or models' explainable outputs) of AI practitioners in addressing responsible AI issues. This strand of research covers various aspects of responsible AI, including fairness, explainability, sustainability, and best practices for data and model documentation and evaluation. Fairness is a fundamental value in responsible AI, but its definition is complex and multifaceted~\cite{narayanan21fairness}. To assess bias in classification outputs, various research efforts have introduced quantitative metrics such as disparate impact and equalized odds, as discussed by Dixon \emph{et al.}~\cite{dixon2018measuring}. Another concept explored in the literature is ``equality of opportunity,'' advocated by Hardt \emph{et al.}\cite{hardt2016equality}, which ensures that predictive models are equally accurate across different groups defined by protected attributes like race or gender. Equally important is the development of dedicated checklists for fairness~\cite{madaio2020co}. Explainable AI (XAI) is another aspect of responsible AI. XAI involves tools and frameworks that assist end users and stakeholders in understanding and interpreting predictions made by machine learning models~\cite{arrieta2020explainable, kulesza2015principles, gunning2019xai,liao2021human, ehsan2020human, ibm2019_XAI}. Furthermore, the environmental impact of training AI models should also be considered. Numerous reports have highlighted the significant carbon footprint associated with deep learning and large language models~\cite{sharir2020cost, hao2019training, strubell2019energy}. Best practices for data documentation and model evaluation have also been developed to promote fairness in AI systems. Gebru \emph{et al.}~\cite{gebru2021datasheets} proposed ``Datasheets for Datasets'' as a comprehensive means of providing information about a dataset, including data provenance, key characteristics, relevant regulations, test results, and potential biases. Similarly, Bender \emph{et al.}\cite{bender2018data} introduced ``data statements'' as qualitative summaries that offer crucial context about a dataset's population, aiding in identifying biases and understanding generalizability. For model evaluation, Mitchell \emph{et al.}~\cite{mitchell2019model} suggested the use of model cards, which provide standardized information about machine learning models, including their intended use, performance metrics, potential biases, and data limitations. Transparent reporting practices, such as the TRIPOD statement by Collins \emph{et al.}~\cite{collins2015transparent} in the medical domain, emphasize standardized and comprehensive reporting to enhance credibility and reproducibility of AI prediction models.

\subsection{Compiling an Initial Catalog of Responsible AI Guidelines}
\label{subsec:step2}
For each research article previously identified, we compiled a list of techniques that could be employed to create responsible AI guidelines, focusing on the actions different roles (i.e., designers, researchers, developers, product managers) should consider during AI development. Following the methodology proposed by Michie \emph{et al.}~\cite{michie2013behavior} (which was also used to identify community engagement techniques by Dittus \emph{et al.}~\cite{dittus2017community}), we sought techniques that describe the ``active ingredient'' of what needs to be done. This means that the phrasing of the technique should focus on \emph{what} needs to  be done, rather than the specific implementation details of \emph{how} it should be done. For example, a recommended practice for ensuring fairness involves evaluating an AI system across different demographic groups~\cite{madaio2020co, dixon2018measuring, hardt2016equality}. In this case, the technique specifies ``what'' needs to be done  (e.g., using common fairness metrics such as demographic parity or equalized odds) rather than ``how'' it should be implemented. In total, we formulated a set of 16 techniques based on relevant literature sources~\cite{mitchell2019model, madaio2020co, dixon2018measuring, hardt2016equality, mitchell2018prediction, verma2018fairness, arrieta2020explainable, kulesza2015principles, fjeld2020principled, bender2018data, gebru2021datasheets, holland2018dataset, wang2020revise, collins2015transparent, sharir2020cost, hao2019training}. 

We then conducted an iterative review of the collection of techniques to identify duplicates, which were instances where multiple sources referred to the same technique. For example, four sources indicated that data biases could affect the model~\cite{mitchell2019model, gebru2021datasheets, bender2018data, holland2018dataset}, emphasizing the need to report the characteristics of training and testing datasets. We consolidated such instances by retaining the specific actions to be taken (e.g., \emph{reporting} dataset characteristics). This process resulted in an initial list of 16 distinct techniques. We provided a concise summary sentence for each technique, utilizing active verbs to emphasize the recommended actions.

\subsection{Refining the Catalog Through Interviews and Expert Panels}
\label{sec:step_adding_iso}
The catalog of techniques underwent eleven iterations to ensure clarity and comprehensive thematic coverage. The iterations were carried out by two authors, with the first author conducting interviews with five AI researchers and developers. During the interviews, the participants were asked to consider their current AI projects and provide insights on the implementation of each technique, focusing on the ``how'' aspect. This served two purposes: firstly, to identify any statements that were unclear or vague, prompting suggestions for alternative phrasing; and secondly, to expand the catalog further. The interviews yielded two main recommendations for improvement: \emph{(1)} mapping duplicate techniques to the same underlying action(s); and \emph{(2)} adding examples to support each technique (each guideline in Table~\ref{tbl:techniques} indeed comes with an example).

In addition to the interviews, the two authors who developed the initial catalog conducted a series of eight 1-hour expert panels with two standardization experts from a large organization. The purpose of these panels was to review the initial catalog for ISO compliance. The standardization experts examined eight AI-related ISOs, including ISO 38507, ISO 23894, ISO 5338, ISO 24028, ISO 24027, ISO 24368, ISO 42001, and ISO 25059, which were developed at the time of writing. Then the experts provided input on any missing techniques and mapped each technique in the initial catalog to the corresponding ISO that covers it. As a result of this exercise, six new techniques (\#2, \#7, \#12, \#13, \#14, \#21 in Table~\ref{tbl:techniques}) were added to the catalog, resulting in a total of 22 guidelines. Next, we provide we provide a high-level summary of each ISO.\footnote{Note that the summary provided is a brief and simplified description due to a paywall restriction.} 
\smallskip

\noindent\textbf{ISO 38507 (Governance, 28 pages).} It offers guidance on responsible AI use (e.g., identify potential harms and risks for each intended use(s) of the systems), and recommendations about current and future AI uses to governing bodies and various stakeholders such as managers and auditors.

\noindent\textbf{ISO 23894 (Risk Management, 26 pages).} It provides guidelines for managing AI-related risks (e.g., mechanisms for incentivizing reporting of system harms) in developing, producing, deploying, or using AI products and systems, including recommendations for integrating risk management into AI processes.

\noindent\textbf{ISO 5338 (AI Lifecycle Process, 27 pages).} It provides a framework for the life cycle of AI systems, detailing processes for managing and enhancing these systems from development to implementation (e.g., through reporting of harms and risks, obtaining approval of intended uses).

\noindent\textbf{ISO 24028 (Trustworthiness, 43 pages).} It offers guidance on trustworthiness in AI systems, focusing on transparency, explainability, controllability, and addressing potential risks with mitigation techniques. It also covers AI systems' availability, resiliency, reliability, accuracy, safety, security, and privacy.

\noindent\textbf{ISO 24027 (Bias, 39 pages).} It discusses bias in AI systems related to protected attributes such as age and gender, especially in AI-aided decision-making, providing techniques to measure and assess bias throughout the AI system lifecycle.

\noindent\textbf{ISO 24368 (Ethical and Societal Concerns, 48 pages).} It provides an introduction to ethical and societal concerns related to AI (e.g., principles, processes, and methods), targeting technologists, regulators, interest groups, and society as a whole.

% \todo{add description of the new ISO 42001 and 25059.}

\noindent\textbf{ISO 42001 (AI Management System, 51 pages).} It outlines the requirements for the establishment, implementation, maintenance, and continuous improvement of an Artificial Intelligence Management System in organizations.

% 1, 

% published: 42001, 25059
% draft: 42005, 42006 (auditing - under commeting period) and EU standards  CEN/CENELEC

% 42005 - AI Impact Assessment - send the draft
% European body to make standards from legislation

\noindent\textbf{ISO 25059 (Quality Model for AI Systems, 15 pages).} It describes characteristics and sub-characteristics that offer a unified terminology for specifying, measuring, and evaluating the quality of AI systems.

As the final step of refining the catalog, the two authors reviewed the 85 articles of the EU AI Act~\cite{eu_ai_act_2022} to map each of the 22 guidelines with the most relevant article(s), as shown in the last column of Table~\ref{tbl:techniques}. They began with Article 3 of the Act, which defines the key concepts of an AI system, including its definition, intended purpose, performance, training, validation, and post-deployment monitoring. After reading all the articles and annotating them, they identified 22 unique articles corresponding to the guidelines. Articles 9, 10, and 17 were mapped to multiple guidelines.
For example, Article 9 (\emph{Risk management system}) states that ``a risk management system shall be established, implemented, documented and maintained throughout the entire lifecycle of a high-risk AI system''. This article aligns with guidelines \#1, \#3-5, and \#13 as it is about the identification of harms and risks of the AI system's intended use. Article 10 (\emph{Data and data governance}) states that \emph{``training, validation and testing data sets shall be subject to appropriate data governance and management practices''.} This article aligns with guidelines \#8 and \#15-18 as it discusses the management and quality of data for training, validation, and testing, including aspects of diversity and minimizing biases. Finally, Article 17 (\emph{Quality management system}) states that \emph{``an AI system shall be documented in a systematic and orderly manner in the form of written policies, procedures and instructions''}. This article aligns with guidelines \#6, \#7, \#10, and \#14-18 because it is about documentation of all system components, including AI models and testing and validation procedures. The full mapping along with justifications is provided in Appendix~\ref{app:mapping_guidelines}.

\begin{table}
\caption{Responsible AI guidelines are actionable items that can be considered during the 3 phases of AI development lifecycle. These guidelines are grounded in the scientific literature (main sources are reported), and were checked for ISO ``compliance'': ISO 38507 (Governance); ISO 23894 (Risk management); ISO 5338 (AI lifecycle processes); ISO 24028 (Trustworthiness); ISO 24027 (Bias); ISO 24368 (Ethical considerations); ISO 42001 (AI management system); and ISO 25059 (Quality model for AI systems). They were also cross-referenced with the EU AI Act's articles~\cite{eu_ai_act_2022}. They are marked with the `Phase' during which a guideline can be applied. There are three phases: development ($P_1$), deployment ($P_2$), and use ($P_3$). Guidelines are also marked with the job `Role' that should consider them. There are three roles: designer ($R_{D}$),  engineer or researcher ($R_E$), and manager or executive ($R_M$). Each guideline is followed by an example, and the guidelines are categorized thematically into six categories, concerning the \emph{intended uses, harms, system, data, oversight, and team}.}
\label{tbl:techniques}
\resizebox{\textwidth}{!}{%
\begin{tabular}{lllllll}
\toprule
\textbf{Number} & \textbf{Guideline} & \textbf{Phase} & \textbf{Role} & \textbf{Source(s)} & \textbf{ISO} & \textbf{AI Act}\\ \midrule
\multicolumn{2}{l}{\textbf{INTENDED USES}} &  &  \\
1 & \begin{tabular}[t]{l}Work with relevant parties to identify intended uses. \\ (e.g., identify the system's usage, deployment, and contextual conditions)\end{tabular} & 
$P_{1-3}$ & $R_{D,E,M}$ &
\begin{tabular}[t]{l} \cite{mitchell2019model} \end{tabular}
 & \begin{tabular}[t]{l} 5338, 38507, 23894, \\ 24027, 24368, 42001\end{tabular} & Art. 6, 9\\

2 & \begin{tabular}[t]{l}Obtain approval from an Ethics Committee or similar body for intended uses.\\ (e.g., Obtain Ethics Committee approval for the intended use, aligned with sustainability goals)\end{tabular} &
$P_{1-3}$ & $R_{D,E,M}$ &
\begin{tabular}[t]{l} ---\end{tabular} & \begin{tabular}[t]{l}38507, 5338, 23894, \\ 42001\end{tabular} & Art. 11, 69 \\ 

\midrule
\multicolumn{2}{l}{\textbf{HARMS}} &  &  \\
3 & \begin{tabular}[t]{l}Identify potential harms and risks associated with the intended uses. \\  (e.g., prevent privacy violation, discrimination, and adversarial attacks, provide interpretable output)\end{tabular} & 
$P_{1-3}$ & $R_{D,E,M}$ &
\begin{tabular}[t]{l} \cite{madaio2020co} \end{tabular} & \begin{tabular}[t]{l}23894, 24028, 38507, \\ 24368, 42001, 25059\end{tabular}  & Art. 9, 65\\

4 & \begin{tabular}[t]{l}Provide mechanism(s) for incentivizing reporting of system harms. \\  (e.g., provide contact emails and feedback form to raise concerns)\end{tabular} &
$P_{1}$ & $R_{D,E}$ &
\begin{tabular}[t]{l} \cite{madaio2020co} \end{tabular} & \begin{tabular}[t]{l}38507, 23894, 42001\end{tabular} & Art. 9, 60-63\\

5 & \begin{tabular}[t]{l}Develop strategies to mitigate identified harms or risks for each intended use. \\ (e.g., use stratified sampling and safeguards against adversarial attacks during training)\end{tabular} &
$P_{1-3}$ & $R_{D,M}$ &
\begin{tabular}[t]{l} \cite{mitchell2019model} \end{tabular}
 & \begin{tabular}[t]{l}24368, 23894,\\ 42001, 25059\end{tabular} & Art. 9, 67\\ 

\midrule
\multicolumn{2}{l}{\textbf{SYSTEM}} &  &  \\
6 & \begin{tabular}[t]{l}Document all system components, including the AI models, to enable reproducibility and scrutiny. \\ (e.g., create UML diagrams, flowcharts, and specify model types, versions, hardware architecture)\end{tabular} &
$P_{1-3}$ & $R_{D,E}$ &
 \begin{tabular}[t]{l} \cite{madaio2020co} \end{tabular} & \begin{tabular}[t]{l}5338, 23894, 24027, \\ 42001, 25059\end{tabular} & Art. 11, 12, 16-18, 50\\

7 & \begin{tabular}[t]{l}Review the code for reliability\\ (e.g., manage version control using software.)\end{tabular} &
$P_{1-3}$ & $R_{D,E}$ &
 \begin{tabular}[t]{l} --- \end{tabular} & \begin{tabular}[t]{l}5338, 25059 \end{tabular} & Art. 17\\

8 & \begin{tabular}[t]{l}Report evaluation metrics for various groups based on factors such as age, gender, and ethnicity. \\ (e.g., evaluate false positive/negative, AUC, and feature importance across protected attributes)\end{tabular} &
$P_{1-3}$ & $R_{D,E,M}$ &
\begin{tabular}[t]{l} \cite{madaio2020co, dixon2018measuring, hardt2016equality} \\ \cite{mitchell2018prediction, verma2018fairness}\end{tabular} & \begin{tabular}[t]{l}23894, 5338, 24028, \\ 24027, 42001 \end{tabular} & Art. 10, 13\\

9 & \begin{tabular}[t]{l} Provide mechanisms for interpretable outputs and auditing. \\ (e.g., output feature importance and provide human-understandable explanations)\end{tabular} &
$P_{1-3}$ & $R_{D,E,M}$ &
 \begin{tabular}[t]{l} \cite{arrieta2020explainable, kulesza2015principles}  \end{tabular} & \begin{tabular}[t]{l}38507, 24028, \\ 42001, 25059\end{tabular} & Art. 12-14\\

10 & \begin{tabular}[t]{l}Document the security of all system components in consultation with experts. \\ (e.g., guard against adversarial attacks and unauthorized access)\end{tabular} &
$P_{1-3}$ & $R_{E,M}$ &
 \begin{tabular}[t]{l} \cite{fjeld2020principled} \end{tabular} & \begin{tabular}[t]{l}24028, 24368, 42001, 25059 \end{tabular} & Art. 12, 13, 15, 17\\

11 & \begin{tabular}[t]{l}Provide an environmental assessment of the system. \\ (e.g., report the number of GPU hours used in training and deployment)\end{tabular} &
$P_{1-3}$ & $R_{E}$ &
\begin{tabular}[t]{l} \cite{sharir2020cost, hao2019training} \end{tabular}
 & \begin{tabular}[t]{l}38507, 23894, 5338, \\ 24368, 42001, 25059\end{tabular} & Art. 69\\

12 & \begin{tabular}[t]{l}Develop feedback mechanisms to update the system. \\ (e.g., provide contact email, feedback form, and notification of new knowledge extracted)\end{tabular} &
$P_{1-3}$ & $R_{D,E}$ &
 \begin{tabular}[t]{l} ---\end{tabular} & \begin{tabular}[t]{l}24028, 42001\end{tabular} & Art. 61\\

13 & \begin{tabular}[t]{l}Ensure safe system decommissioning.\\ (e.g., ensure decommissioned data is either deleted or restricted to authorized personnel.)\end{tabular} &
$P_{3}$ & $R_{E}$ &
 \begin{tabular}[t]{l} ---\end{tabular} & \begin{tabular}[t]{l}38507, 24368, 42001\end{tabular} & Art. 9\\

14 & \begin{tabular}[t]{l}Redocument model information and contractual requirements at every system update.\\  (e.g., update the model information when re-training the system or using datasets with new contractual requirements)\end{tabular} &
$P_{3}$ & $R_{E}$ &
 \begin{tabular}[t]{l} ---\end{tabular} & \begin{tabular}[t]{l}23894, 5338, 24368, \\ 42001 \end{tabular} & Art. 11, 12, 17, 61\\ 

\midrule
\multicolumn{2}{l}{\textbf{DATA}} &  &  \\

15 & \begin{tabular}[t]{l}Ensure compliance with agreements and legal requirements when handling data. \\  (e.g., create data sharing and non-disclosure agreements and secure servers)\end{tabular} &
$P_{1-3}$ & $R_{D,E,M}$ &
 \begin{tabular}[t]{l} --- \end{tabular} & \begin{tabular}[t]{l}38507, 23894, 5338 \\ 42001\end{tabular} & Art. 10, 17, 61\\ 

16 & \begin{tabular}[t]{l}Compare the quality, representativeness, and fit of training and testing datasets with the intended uses. \\  (e.g., report dataset details such as public/private, personal information, demographics, and data provenance)\end{tabular} &
$P_{1-3}$ & $R_{E}$ &
 \begin{tabular}[t]{l}\cite{bender2018data, gebru2021datasheets, holland2018dataset, wang2020revise}\\ \cite{madaio2020co, mitchell2018prediction, verma2018fairness}\end{tabular}
& \begin{tabular}[t]{l}38507, 5338, 24028, \\ 24027, 42001, 25059\end{tabular} & Art. 10, 13, 17, 64 \\

17 & \begin{tabular}[t]{l}Identify any measurement errors in input data and their associated assumptions. \\ (e.g., account for potential input errors in the input device, text data, audio, and video)\end{tabular} &
$P_{1-3}$ & $R_{E}$ &
\begin{tabular}[t]{l} \cite{collins2015transparent} \end{tabular} & \begin{tabular}[t]{l}38507, 42001, 25059\end{tabular} & Art. 10, 13, 17, 64 \\

18 & \begin{tabular}[t]{l}Protect sensitive variables in training/testing datasets. \\ (e.g., protect sensitive data and use techniques such as k-anonymity and differential privacy)\end{tabular} &
$P_{1-3}$ & $R_{D,E,M}$ &
\begin{tabular}[t]{l} \cite{dworkdifferential} \end{tabular} & \begin{tabular}[t]{l}38507, 24028, 42001\end{tabular} & Art. 10, 13, 17 \\ 

\midrule
\multicolumn{2}{l}{\textbf{OVERSIGHT}} &  &  \\
19 & \begin{tabular}[t]{l}Continuously monitor metrics and utilize guardrails or rollbacks to ensure the system's output stays within a desired range. \\ (e.g., validate against concept drift and test with diverse testers and compliance and adversarial cases)\end{tabular} &
$P_{1-3}$ & $R_{D,E}$ &
\begin{tabular}[t]{l} \cite{fjeld2020principled} \end{tabular}  & \begin{tabular}[t]{l}38507, 5338, 24028, \\ 24027, 24368, 42001\end{tabular} & Art. 12, 20, 29, 61 \\

20 & \begin{tabular}[t]{l}Ensure human control over the system, particularly for designers, developers, and end-users.  \\ (e.g., include human in the loop with the ability to inspect data, models, and training methods)\end{tabular} &
$P_{1-3}$ & $R_{D,E,M}$ &
\begin{tabular}[t]{l} ---\end{tabular} & \begin{tabular}[t]{l}38507, 5338,\\ 24028, 24368, 25059\end{tabular} & Art. 13, 14\\ 

\midrule
\multicolumn{2}{l}{\textbf{TEAM}} &  &  \\

21 & \begin{tabular}[t]{l}Ensure team diversity.  \\ (e.g., consider diversity in gender, neurotypes, personality traits, and thinking styles)\end{tabular} &
$P_{1-3}$ & $R_{D,E,M}$ &
\begin{tabular}[t]{l} --- \end{tabular} & \begin{tabular}[t]{l}38507, 5338, 24028, \\ 24368, 42001\end{tabular} & Art. 69\\ 

22 & \begin{tabular}[t]{l}Train team members on ethical values and regulations. \\ (e.g., train on privacy regulations, ethical issues, and raising concerns)\end{tabular} &
$P_{1-3}$ & $R_{D,E,M}$ &
\begin{tabular}[t]{l} \cite{fjeld2020principled}\end{tabular}  & \begin{tabular}[t]{l}38507, 24368, 42001\end{tabular} & Art. 69\\ 

\bottomrule
\end{tabular}%
}
\end{table}

\subsection{Finalizing the Catalog}
In response to the interviews with AI developers and standardization experts, we incorporated an example for each guideline. For instance, under the guideline on system interpretability (guideline \#9), the example provided reads: ``output feature importance and provide human-understandable explanations.'' Furthermore, we simplified the language by avoiding domain-specific or technical jargon. We also categorized each guideline into six thematically distinct categories, namely \emph{intended uses}, \emph{harms}, \emph{system}, \emph{data}, \emph{oversight}, and \emph{team}.

Recognizing that certain guidelines may only be applicable at specific stages (e.g., monitoring AI after deployment) by specific roles (e.g., developers, managers), we went through two steps. First, we assigned the guidelines to three phases based on previous research (e.g., \cite{madaio2020co, mitchell2019model}). These phases are development (designing and coding the system), deployment (transferring the system into the production stage), and use (actual usage of the system). For example, guidelines like identifying the system's intended uses (guideline \#1) are relevant to all three phases, while those related to system updates (guideline \#14) or decommissioning (guideline \#13) are applicable during the use phase. 

Second, based on previous literature, we assigned the guidelines to the three roles of designers, engineers/researchers, and managers/executives  (Table~\ref{tbl:techniques}). Wang \emph{et al.}~\cite{wang2023designing} interviewed UX practitioners and responsible AI experts to understand their work practices. UX practitioners included designers, researchers, and engineers, while responsible AI experts included ethics advisors and specialists. Wong \emph{et al.}~\cite{wong2023seeing} analyzed 27 ethics toolkits to identify the intended audience of these toolkits, specifically those who are expected to engage in AI ethics work. The intended audience roles identified included software engineers, data scientists, designers, members of cross-functional or cross-disciplinary teams, risk or internal governance teams, C-level executives, and board members. Additionally, Madaio \emph{et al.}~\cite{madaio2020co} co-designed a fairness checklist with a diverse set of stakeholders, including product managers, data scientists and AI/ML engineers, designers, software engineers, researchers, and consultants. Following guidance  from these studies~\cite{wang2023designing,wong2023seeing,madaio2020co}, we formulated three roles as follows: 
\begin{enumerate}
    \item Designer: This role includes interaction designers and UX designers.
    \item Engineer or Researcher: This role includes AI/ML engineers, AI/ML researchers, data scientists, software engineers, UX engineers, and UX researchers.
    \item Manager or Executive: This role includes  product managers, C-suite executives, ethics advisors/responsible AI consultants, and ethical board members.
\end{enumerate} 

The revised and final catalog, consisting of 22 unique guidelines, is presented in Table~\ref{tbl:techniques}.

\section{Evaluation of the 22 Responsible AI Guidelines}
\label{sec:userstudy}

We first conducted a formative study with 10 AI practitioners from a large technology company to elicit design requirements for an evaluation tool, implemented the tool (Panel B in Figure~\ref{fig:card-elements} and \S\ref{subsec:populate_tool}), and relied on it to conduct a user study with 14 other AI researchers, engineers, designers, and product managers from the same company (Panel C in Figure~\ref{fig:card-elements} and \S \ref{subsec:evaluate_guidelines}).

\subsection{Incorporating  the guidelines into a tool} 
\label{subsec:populate_tool}

\mbox{ } \\
\noindent\textbf{Eliciting design requirements for a tool through a formative study.} We conducted a formative study that included semi-structured interviews with 10 participants. These participants, comprising 6 males and 4 females, were AI practitioners in their 30s and 40s employed at a large technology company. The participants had a range of work experience, spanning from 1 to 8 years, and were skilled in areas such as data science, data visualization, UX design, natural language processing, and machine learning. The interview study took place online and consisted of three parts. In the first part, we encouraged participants to share information about their ongoing AI projects. In the second part, we presented them with the table containing the 22 guidelines and asked them to think about how each guideline could apply to their projects. Finally, in the third part, we conducted semi-structured interviews to discuss how these guidelines could be incorporated into an interactive responsible AI tool.
 
Each study lasted about half an hour. Two authors took notes during the interviews, and afterward, they analyzed the interview transcripts using inductive thematic analysis~\cite{saldana2015coding, miles1994qualitative, mcdonald2019reliability, braun2006thematic}. This analysis then resulted in the following four design requirements (participant quotes are marked with FP):
\smallskip

\noindent
\emph{R1: Simplify the guidelines by breaking them into smaller visual components.} Participants found it challenging to reflect on guidelines and examples because of their quantity. According to FP5, \emph{``the sheer number of the guidelines is the main difficulty [...] they should be separated in bite-sized questions''}. Additionally, participants requested to visually separate the guidelines from the examples.
\smallskip

\noindent
\emph{R2: Implement clear navigation features to systematically guide users through the guidelines.} 
Participants were unsure about the best way to navigate through the guidelines. FP9 suggested that \emph{``the system should provide clear navigation [...] for example, using a progress bar''}. FP5 further emphasized that the design of the progress bar could facilitate \emph{``gaining insights while engaging with the 22 guidelines''}.
\smallskip

\noindent
\emph{R3: Track how guidelines are applied and share progress among team members.} 
Participants faced difficulty in tracking their responses on how to apply the guidelines to their projects and share progress among team members. To address this challenge, FP5 suggested implementing a feature that would save user responses as they progress through the guidelines: \emph{``there should be some functionality there that captures the answers I gave, so it'd allow me to track progress and share it among team members''}. These responses would then be transformed into comprehensive documentation and made accessible to users for download.
\smallskip

\noindent
\emph{R4: Develop a mechanism for post-hoc reflections on how the project aligns with responsible AI guidelines.} 
Participants found it challenging to envision how well their AI systems aligned with the guidelines. Therefore, FP8 suggested developing \emph{``visual feedback or a score that shows how responsible [their] AI system is.''} However, FP2 cautioned that this mechanism \emph{``should not make me anxious and feel like I have not done enough''}. Instead, it should create a positive learning experience and encourage users to generate ideas for improving their AI systems.
\smallskip

\begin{figure*}[ht!]
  \includegraphics[width=0.75\textwidth]{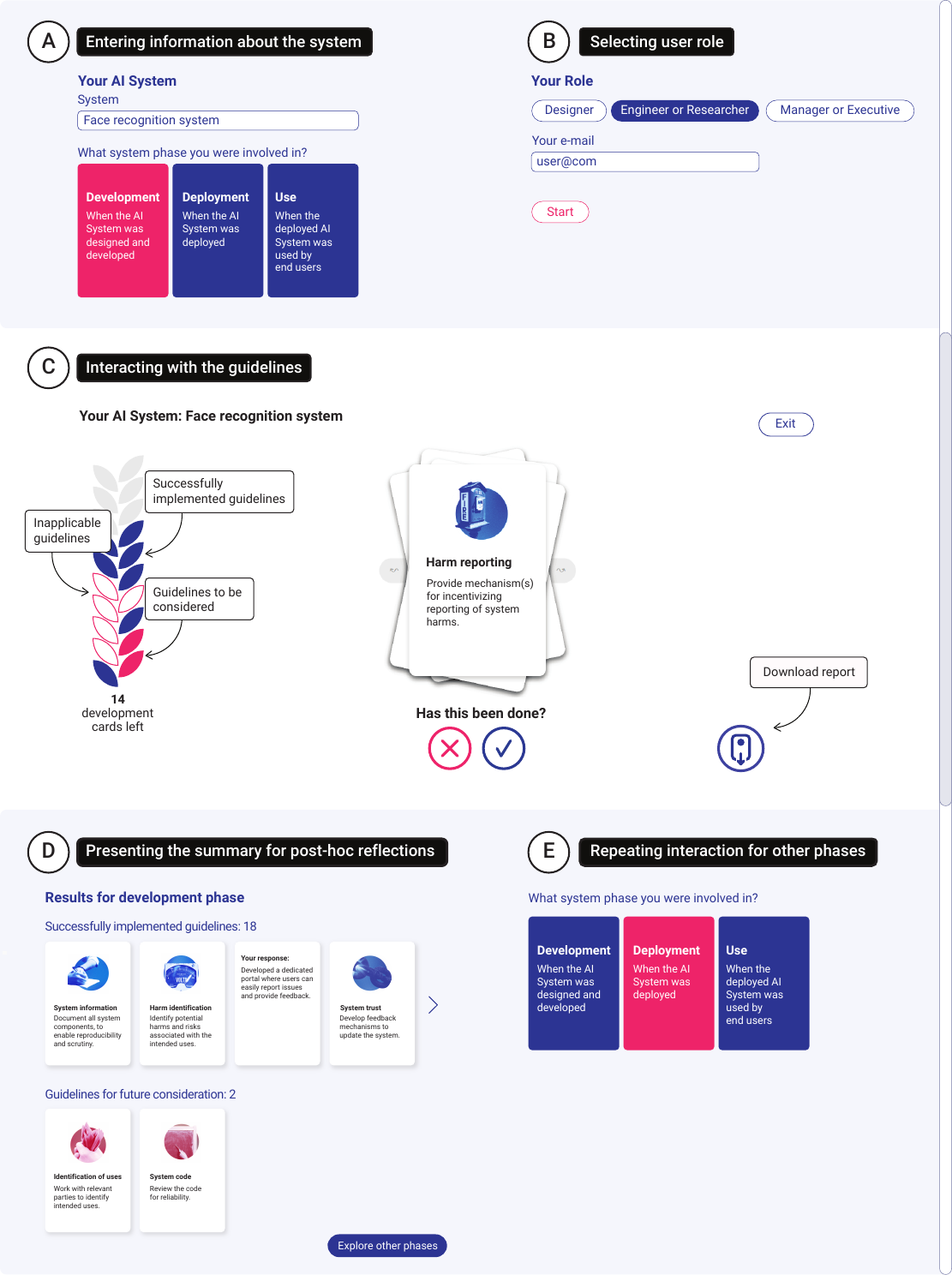}
  \caption{Interactive Responsible AI Tool with 22 guidelines. The first part (A) allows for entering information about the developed AI system and (B) selecting the applicable user role. The second part (C) enables interaction with the guidelines. The third part (D) presents a summary of user responses for post-hoc reflections. Guidelines for other project phase can be viewed through the phase selectors (E/A).}
  \label{fig:ui-sections}
\end{figure*}

\mbox{ } \\
\noindent\textbf{Designing the tool and incorporating the guidelines.} To meet these requirements, we designed an interactive web-based tool\footnote{\url{https://social-dynamics.net/rai-guidelines}} (Figure~\ref{fig:ui-sections}) and populated it with the 22 guidelines in Table~\ref{tbl:techniques}.

To meet  design requirement R1 (\emph{Simplify the guidelines}), each guideline is presented as a digital card \cite{RAIPatterns} with interactive boxes on both the front and back sides. The front side includes a symbolic graphic collage representing the guideline, followed by its name and full text. The back side includes an input box for users to write their thoughts on implementing each guideline in their project \cite{sanderson2023ai}. We also used this box to showcase an example for each guideline (refer to Figure~\ref{fig:card-elements}). Initially, the example in the box is visible, but it disappears once the user inputs their specific implementation details. Users can view the guideline from both sides by using the flip buttons at the bottom-left corner of each side.

\begin{figure*}[t!]
  \includegraphics[width=\textwidth]{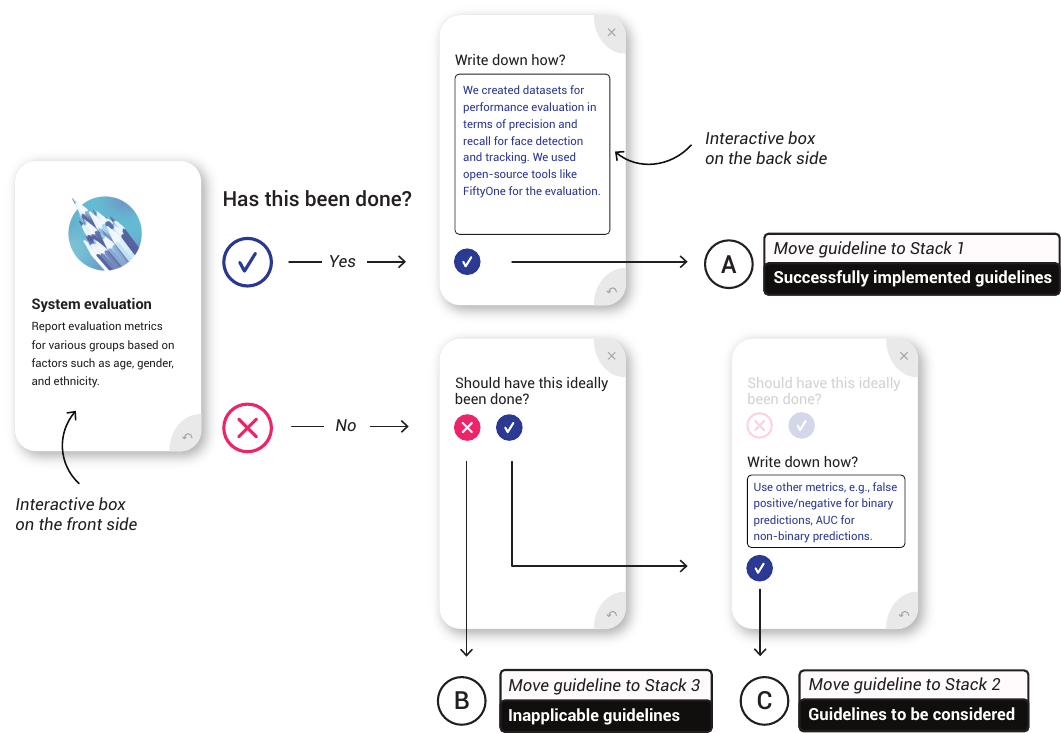}
  \caption{Guideline sorting procedure. Users can place a guideline in any of the three stacks (i.e., successfully implemented, should be considered, inapplicable) by: (1) considering two guiding questions, and (2) using the Yes/No buttons located next to the card and on the back side of it.}
  \label{fig:game-sorting}
\end{figure*}

Each guideline is paired with two guiding questions~\cite{yildirim2023investigating} that help users think about the relevance of the guideline to their specific AI system and context (Figure~\ref{fig:game-sorting}). The first question asks the user whether the guideline has been successfully implemented in their AI system. For example, for an engineer addressing fairness, the question asks if they have reported evaluation metrics for various groups based on factors like age, gender, and ethnicity (technique \#8 in Table~\ref{tbl:techniques}). If the engineer answers ``yes'', they are then prompted to provide specific details on how fairness was implemented in the input box on the card's back. After sharing this information, the tool moves the guideline to the ``successfully implemented'' stack. In contrast, if the engineer answers ``no'', the tool asks a second follow-up question regarding whether the guideline should be implemented in a future iteration. If the engineer answers ``yes'', they are prompted to provide specific details on how to implement it. The tool then moves the guideline to the ``should be considered'' stack. However, if the engineer answers ``no'' to both questions, indicating that the guideline is not applicable to their AI system, the tool moves the guideline to the ``inapplicable'' stack.

\smallskip

To meet design requirement R2 (\emph{Implement clear navigation}), we explored different layout options and considered previous research that involved swiping ~\cite{cardInteractions_2020}, scrolling, or organizing guidelines into different groups ~\cite{dittus2017community}. Due to the limited screen size and the repetition of guidelines for each phase and role, we chose to organize the guidelines into nine groups. These groups were derived from three phases of the AI system: development (designing and coding),  deployment (transitioning into production), and use (actual usage of the system), as well as from three user roles:  designer,  engineer or researcher, and  manager or executive. The number of guidelines in each group varied and accommodated the specific requirements of each phase and role. For example, engineers or researchers needed to go through 20 guidelines for development, 18 for deployment, and 20 for use (\S\ref{sec:method}, Step 4).

%To facilitate the browsing of guidelines and addressing them in the user's preferred order, we introduced two arrow buttons on both the left and right side of the card group. We then introduced a graphical progress bar next to the group that not only displayed the number of remaining guidelines but also color-coded them to indicate their assignment to the three stacks. Blue leaves in the bar represented successfully implemented guidelines, magenta leaves represented guidelines for future consideration, and empty leaves represented inapplicable guidelines. Finally, we added an ``Exit'' button that becomes available as soon as the user goes through minimum two guidelines. In that way user can quit the experience at a preferable moment.

To meet design requirement R3 (\emph{Track how guidelines are applied and share progress among team members}), we added a feature to store user responses locally in the browser session. Users can download their responses as a structured PDF report at any time.

To fulfill the last requirement, R4 (\emph{Develop a mechanism for post-hoc reflections}), after completing the sorting process, we display a summary page to the user. The summary is divided into three sections, one for each stack of cards (i.e., successfully implemented, should be considered, and inapplicable), with in-text counters indicating the number of guidelines in each stack. To read the responses for each guideline, hover-over functionality is provided.

Figure~\ref{fig:ui-sections} shows the tool with its three parts that meet the four design requirements. The first part enables users to enter the name of the developed AI system (Figure \ref{fig:ui-sections}A), select the phase it belongs to and specify the user's role (Figure \ref{fig:ui-sections}B). Once the phase and role are selected, the second part displays the guidelines one by one (Figure \ref{fig:ui-sections}C). The third part presents the user with the summary for post-hoc reflections (Figure \ref{fig:ui-sections}D). If desired, the user can repeat the experience and generate documentation for other phases (Figure \ref{fig:ui-sections}E).

\subsection{Evaluating the Guidelines Through a User Study}
\label{subsec:evaluate_guidelines}
To evaluate whether our guidelines are usable by different roles and whether they match the EU AI Act articles and ISO standards, we conducted a user study with 14 AI researchers, engineers, designers, and managers (Panel C in Figure~\ref{fig:card-elements}).
\smallskip

\noindent\textbf{Participants.}
 The recruitment process took place in October and November 2022.\footnote{Participants who took part in the formative study were not eligible to participate in this evaluation study.} We aimed for a balanced sample of participants, including a variety of roles such as researchers (5), designers (3), engineers (3), and managers (3). All participants had significant expertise in AI, including areas such as machine learning, deep learning, and computer vision. Additionally, each participant was actively involved in at least one ongoing AI project during the time of the interviews. Table~\ref{tab:demographics} summarizes participants' demographics.
\smallskip

\begin{table*}[t!]
    \centering
    \caption{User study participants' demographics, including their job `Role' (designer ($R_{D}$),  engineer or researcher ($R_E$), and manager or executive ($R_M$)).}
    \label{tab:demographics}
    \resizebox{\textwidth}{!}{%
    \begin{tabular}{lllllll}
    \toprule
    \textbf{ID} & \textbf{Gender} & \textbf{Yrs of expr. in AI} & \textbf{Education} & \textbf{Current continent} & \textbf{Expertise} & \textbf{Role}\\ \midrule
    1 & Male & 6 & Ph.D. & EU & Deep learning, computer vision & $R_M$ \\
    2 & Male & 10+ & Ph.D. & North America & Machine learning, computer vision & $R_E$ \\
    3 & Male & 8 & Ph.D. & EU & Machine learning & $R_E$ \\
    4 & Male & 4 & Ph.D. & North America & Deep learning, IoT, computer vision & $R_E$ \\
    5 & Female & 5 & Ph.D. & EU & Machine learning & $R_{D}$ \\
    6 & Female & 8 & Ph.D. & EU & Computer vision & $R_{D}$ \\
    7 & Male & 2 & Ph.D. & North America & Computer vision & $R_E$ \\
    8 & Male & 10 & Ph.D. & EU & Machine learning & $R_M$ \\
    9 & Male & 4 & Ph.D. & North America & Computer vision & $R_E$ \\
    10 & Male & 10+ & M.Sc. & EU & Machine learning, natural language processing & $R_E$ \\
    11 & Male & 10+ & Ph.D. & EU & Machine learning & $R_M$  \\
    12 & Male & 6 & Ph.D. & EU & Machine learning & $R_E$ \\
    13 & Male & 4 & Ph.D. & EU & Reinforcement learning, decision making & $R_E$ \\
    14 & Male & 8 & Ph.D. & EU & Computer vision, robotics & $R_{D}$  \\ \bottomrule
    \end{tabular}%
    }
\end{table*}

\noindent\textbf{Procedure.}
Ahead of the interviews, we sent an email to all participants, providing a concise explanation of the study along with a brief demographics survey. The survey consisted of questions regarding participants' age, domain of expertise, role, and years of experience in AI system development. The survey is available in Appendix~\ref{app:demographics-survey}. It is important to note that our organization, Nokia Bell Labs, approved the study, and we adhered to established guidelines for user studies, ensuring that no personal identifiers were collected, personal information was removed, and the data remained accessible solely to the research team.

During the interview session, we presented either of these two systems to the participants: (1) our tool with the 22 guidelines; or (2) a web page with the checklist items from Microsoft's Fairness Checklist. We used the Microsoft's AI Fairness Checklist as a baseline alternative because it is a published work in a human-computer interaction conference (CHI 2020), is freely available, and has a rigorous, transparent creation process.\footnote{We decided not to compare our tool with existing card-based systems for responsible AI as they serve different purposes. Card-based systems such as the IDEO AI Ethics and the Feminist Tech card aim at providing thought-provoking activities~\cite{IDEODeck_2019} and stimulating ethical conversations~\cite{FeministDeck_2022}. However, they cannot be used as tools ensuring compliance with internal ethical procedures (like Microsoft's AI Fairness Checklist) or ISO standards.} We asked participants to interact with each system for 20 minutes (or less, if finished sooner), alternating between them to avoid any learning effect. To make the scenario as realistic as possible, we encouraged participants to reflect on their ongoing AI projects and consider how the guidelines could be applied in their roles. We also presented them with excerpts from the EU AI Act articles~\cite{eu_ai_act_2022} and summaries of each ISO standard (\S\ref{sec:step_adding_iso}), and asked them whether the guidelines link to these articles and summaries.  We further engaged participants by asking about their preferences, dislikes, and the relevance of the guidelines to their work. Subsequently, we administered the System Usability Scale (SUS)~\cite{brooke1996sus} to assess the usability of the guidelines and the checklist items.

We piloted our study with two researchers (1 female, 1 male), which helped us make minor changes to the study guide (e.g., clarifying question-wording and changing the order of questions for a better interview flow). These pilot interviews were not included in the analysis.
\smallskip

\noindent\textbf{Analysis.} First, we compared the two usability scores after using each system (i.e., the guidelines and the checklist items). Second, two authors conducted an inductive thematic analysis (bottom-up) of the interview transcripts, following established coding methodologies~\cite{saldana2015coding, miles1994qualitative, mcdonald2019reliability}.  The transcripts included  how the guidelines could be applied in the ongoing AI projects, how they link to the EU AI Act articles and ISO standards, and any other preferences or dislikes. The authors used sticky notes on the Miro platform~\cite{miro2022} to capture the participants' answers, and collaboratively created affinity diagrams based on these notes. They held seven meetings, totaling 14 hours, to discuss and resolve any disagreements that arose during the analysis process. Feedback from the last author was sought during these meetings. In some cases, a single note was relevant to multiple themes, leading to overlap between themes. All themes included quotes from at least two participants, indicating that data saturation had been achieved~\cite{guest2006many}. As a result, participant recruitment was concluded after the $14^{\text{th}}$ interview. 
\smallskip

\begin{figure*}[t!]
  \includegraphics[width=0.37\textwidth]{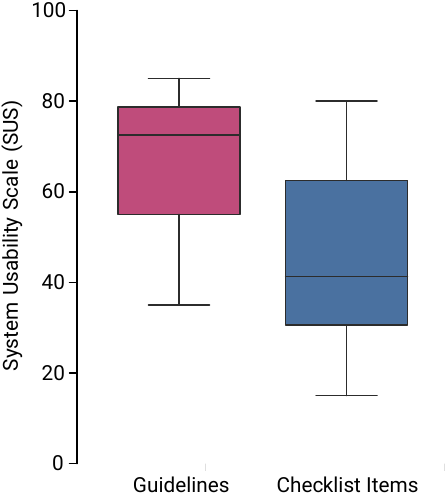}
  \caption{SUS (usability) results. Guidelines are more usable than checklist items.}
  \label{fig:sus_boxplots}
\end{figure*}

\noindent\textbf{Results.}
\label{sec:users-results} Participants, on average, rated the guidelines' usability with a score of 66 out of 100 in SUS, with a standard deviation of 16.01 (Figure~\ref{fig:sus_boxplots}). This indicates a generally positive user experience~\cite{sauro2011practical}. The moderately high usability score was attributed to factors such as familiarity and efficiency in interacting with the guidelines, which were considered usable by different roles. In contrast, participants, on average, rated the checklist items' usability with a score of 44 out of 100 is SUS, with a standard deviation of 21.16. Despite the comparative lower SUS score, checklist items were seen as relevant for audit, formal processes, and certification purposes---acting as a `safeguard'. As for the thematic analysis, the resulting themes are provided in Table~\ref{tab:codebook-interviews} in the Appendix. These themes pertain to how our participants saw the application of guidelines, what worked well, and what could be improved.

\smallskip

Guidelines were generally well-received by the participants. The majority of them (12 out of 14 participants) considered the guidelines valuable for raising awareness and facilitating self-learning about responsible AI, though to different extents.
Participants found the set of guidelines to be comprehensive and aligned with their roles (10 out of 14 participants), as evidenced by P8's observation that \emph{``There are some aspects of responsible AI in the project that I knew about, but I never faced them in such an organized manner''}. Similarly, P4 \emph{``felt that the guidelines were concrete and well-scoped, instead of the lengthy documents of current regulations}.'' Participants also stated that the guidelines align with current regulations (10 out of 14 participants). P7 mentioned that \emph{``he could understand the guidelines relevancy to the ISO standards and their applicability to his work.''} Similarly, P11 found the excerpts from the EU AI Act \emph{``relevant and the guidelines helped him to reflect how the current regulations will affect his project''.} Additionally, seven participants acknowledged the usefulness of the provided examples, which helped them think about potential scenarios and make the guidelines more actionable. One participant expressed that \emph{``the guidelines made me reflect on my previous choices and how I would describe my decisions when I had to develop the system (P3).''} Finally, after becoming familiar with the guidelines, P2 felt more empowered to introduce the topic of responsible system development during group discussions with his team, stating that \emph{``I can at least raise a few questions during team discussions---these are some additional aspects we may need to consider.''}

Participants also offered suggestions for further refinement of the guidelines. Although they found the guidelines aligned with their roles, they expressed the desire for solving team coordination challenges. For example, P6 stated that \emph{``It would be helpful if the guidelines were tailored to the specific challenges I encounter in my project such as seeking feedback from other people''}. P3 specifically mentioned four guidelines about data (listed 15-18 in the Table 1 in the manuscript) and thought of the following improvements: \emph{``I would collect more annotated data from diverse populations and incentivize underrepresented groups to participate in data annotation''.} An action plan was also devised by P1, who recognized that, \emph{``I need an expert in different areas of assessments, because I am probably not in the right position to do that.''} Other participants expressed the need to see how other team members completed the guidelines. For example, P4 stated, \emph{``I want to see how others in a similar role to mine have answered the guidelines''.} In fact,  sharing team members answers in real-time could indeed help reduce the effort required to go through the guidelines. However, this suggestion comes with trade-offs. On one hand, sharing answers among team members not only helps reduce the required effort to go through the guidelines but also helps alleviate the ``blank page syndrome'', also known as ``writer's block''~\cite{bastug2017phenomenological}, which refers to the inability to begin or continue writing due to a lack of ideas, motivation, or confidence. On the other hand, providing team members' answers might hinder an individual's creativity and limit diverse perspectives in the way guidelines are implemented.

\section{Discussion}
\label{sec:discussion}
To assist AI practitioners in navigating the rapidly evolving landscape of AI ethics, governance, and regulations, we have developed a method for generating responsible AI guidelines that are grounded in regulation and usable by different roles. We validated our method in a user study at a large technology company, where we designed and evaluated a tool that incorporates our responsible AI guidelines. We conducted a formative study involving 10 AI practitioners to design the tool, and evaluated our guidelines in a user study with an additional 14 AI practitioners. The results indicate that the guidelines were perceived as practical and actionable, promoting self-reflection and enhancing understanding of the ethical considerations associated with AI during the early stages of development. 

We now discuss the inherent problem of decontextualization in responsible AI toolkits; dwell on the concept of meta-responsibility; and provide practical recommendations for incorporating responsible AI guidelines into toolkits and for  enabling organizational accountability. 

% The inherent challenge in responsible AI toolkits lies in their attempt to reconcile the tension between scalability and context specificity~\cite{wong2023seeing}.

\subsection{Theoretical Implications}

\noindent 
\textbf{Decontextualization.}  Traditional approaches to toolkit development have often favored a universal, top-down approach that assumes a one-size-fits-all solution~\cite{kelty_participatory_toolkit, mattern_toolkit}. However, participatory development, such as the methodology we followed in designing and populating a responsible AI tool with our guidelines, emphasizes the importance of tailoring responsible AI guidelines to specific contexts and job roles needs. Various AI professionals like designers, developers, engineers, and executives have unique needs and concerns. Treating them all the same can lead to issues like decontextualization in responsible AI toolkits~\cite{wong2023seeing}.

To tackle this problem, our proposed method incorporates two key elements: \emph{guidelines usable by different roles} and \emph{guiding questions}. Firstly, the integration of guidelines tailored to different roles and projects provides practical steps and recommendations that technical practitioners can easily implement, or C-level executives can make informed decisions upon. These guidelines serve as a starting point for ethical decision-making throughout the AI lifecycle, contributing to the vision of responsible AI by design (borrowing from the idea of `privacy by design'\footnote{``Privacy by design'' is a standard practice for incorporating data protection into the design of technology. In other words, data protection is achieved when it is already integrated into the technology during its design and development~\cite{cavoukian2009privacy}.}). Secondly, the inclusion of the two guiding questions (\S\ref{subsec:populate_tool}), one on how the guideline was implemented, and the other on how it could have been implemented, enhances our toolkit's ability to capture the complexities of different social and organizational contexts. 
\smallskip

\noindent\textbf{Meta-responsibility.} Scholars have long recognized the need for a socio-technical approach that considers the contextual factors governing the use of AI systems, including social, organizational, and cultural factors~\cite{tahaei2023toward}. In fact, Ackerman~\cite{ackerman2000intellectual} introduced the concept of socio-technical gap to highlight the disparity between human requirements and  technical solutions.  Along similar lines, \citet{stahl2023embedding} introduced the concept of meta-responsibility to stress that AI systems should be viewed as systems of systems  rather than single entities. Our work contributes to the integration of ethical, legal, and social knowledge into the AI development process---what Stahl referred to as ``adaptive governance structure''.

\subsection{Practical Implications}

\noindent \textbf{Recommendations for incorporating responsible AI guidelines into toolkits.} Our work identified four essential design requirements for incorporating guidelines into tools. They include: simplifying guidelines into smaller visual components; implementing clear navigation; tracking and sharing progress; and developing mechanisms for reflection.

For simplifying guidelines, we displayed each guideline as a digital card and accompanied it with two guiding questions. Future work could explore how to further divide guidelines into additional visual elements on the cards and how to refine the guiding questions. For example, guideline \#15---\emph{ensuring compliance with agreements and legal requirements when handling data}---could be further divided into step-by-step processes, with each one marked by a visual element like a card tab or a link to a specific ISO, or excerpts from the EU AI Act. Regarding the guiding questions, we observed that their formulation  is a delicate task, requiring a balance between directness and respect for the user's autonomy. For example, a question formulated as ``How did you consider the potential impact of your AI system on different user groups?'' employs a proactive stance, avoiding any direct accusation or presumption of oversight. This method resonates with the experiences of our participants (e.g., P14) who found value in open-ended questions. However, guiding questions can be refined in various ways by, for example, ``reminding consequences'' or ``providing multiple viewpoints''~\cite{caraban2019ways}.

For ensuring clear navigation, we organized the guidelines into a one-page layout and incorporated multiple buttons along with a counter for easy navigation. Future work could explore how to develop alternative layouts and include different navigation mechanisms. For example, complementary guidelines with related content, such as guideline \#3 \emph{identify potential harms and risks associated with the intended use} and guideline \#5 \emph{develop strategies to mitigate identified harms or risks for each intended use} can be paired side by side to improve the quality of responses. Additionally, new navigation mechanisms might include a chart to illustrate the relationships between guidelines and a search bar to enable users to quickly locate specific guidelines.

For tracking how guidelines are applied and sharing progress among team members, we introduced a feature to store user responses locally within the browser session and dynamically generate a PDF report from these responses. Future work could explore how to structure user responses in formats suitable for automated analysis and integration with other tools. For instance, using JSON format as input for machine learning algorithms and Large Language Models (LLMs) can enable the analysis of user responses and the generation of automated insights and recommendations within the PDF report.

For enabling post-hoc reflections, we created a summary page where users can view the number of guidelines they have considered and their responses to each guideline. Future work could explore how to improve this summary page, for example, by adding visual elements for recognizing responsible AI champions (e.g., responsible AI badges) and fostering empathy (e.g., animations presenting the environmental impact of an AI system), or by implementing a collaborative aspect where users can share and discuss their summary pages with peers or mentors.
\smallskip

\noindent \textbf{Recommendations for enabling organizational accountability.}
While individual adoption of responsible AI best practices is crucial, fostering effective communication between technical and non-technical roles is equally important. Many existing responsible AI toolkits prioritize individual usage~\cite{wong2023seeing}. However, addressing complex ethical and societal challenges associated with AI systems requires diverse perspectives. Our interactive tool populated with guidelines addresses this need by offering features that make the guidelines usable by different roles (e.g., adjusting which guidelines are shown to different roles and in different system phases). However, our tool can further improve communication between roles by creating a knowledge base of responses. Such a knowledge base, according to Stahl \cite{stahl2023embedding}, empowers team members to fulfill their responsibilities and supports distributed teams in constructing a shared understanding of their AI system. Furthermore, we suggest a mechanism for keeping this knowledge base up to date and enriched with diverse perspectives. This includes regularly revisiting the guidelines through our tool and providing responses at key project milestones, such as when the AI system enters a new phase. This approach ensures that the knowledge base remains dynamic and reflect the evolving insights and perspectives within the team.

Our guidelines and the tool that incorporates them can also be used to enable organizational accountability. Similar to Google's five-stage internal algorithmic auditing framework~\cite{raji2020closing}, our guidelines serve as a practical tool for partially closing the AI accountability gap. The automatically generated report plays a crucial role in this process by providing a summary of the guidelines that were effectively implemented, and of those that should be considered for future development. These reports establish an additional chain of accountability that can be shared with stakeholders at various levels, including managers, senior leadership, and AI engineers. By offering more oversight and the ability to troubleshoot, if needed, these reports help mitigate unintentional harm. When an organization follows our guidelines, it needs to set up clear processes though. If incentives are not right, AI professionals may avoid using them because they fear being responsible for their actions.

\subsection{Limitations and Future Work} Our work has four main limitations that highlight the need for future research efforts. Firstly, although we followed a rigorous four-step process involving multiple stakeholders, the list of 22 guidelines may not be exhaustive. The rapidly evolving nature of AI ethics, governance, and regulations necessitates an ongoing effort to stay abreast of emerging developments. However, one of the strengths of our method lies in its modular design, which allows for ongoing refinement and expansion of the set of guidelines.  Future work could incorporate ISOs that are currently under development such as those for functional safety (ISO 5469), data quality (ISO 5259), explainability (ISO 6254), AI system impact assessment (ISO 42005), and requirements for bodies providing audit and certification of AI management systems (ISO 42006). Additionally, the European Committee for Electrotechnical Standardization~\cite{cen_cenelec} (CEN-CENELEC) body was recently tasked to translate the EU AI Act into standards; such standards can also be cross-referenced with our guidelines as part of future work. However, we acknowledge that there may be limitations in ensuring that all standards are accessible to everyone and that experts may not always be available to evaluate them. A partial solution would be to create forums or discussion groups where individuals can share their experiences and insights about regulations and standards. At the same time, future research could also investigate the frequency with which our method should be updated as new literature emerges. One possibility would be to create an automated system that regularly collects research articles on responsible AI best practices, pairing them with current and upcoming regulations, to extract new guidelines.

Secondly, it is important to consider the qualitative nature of our user study. It  involved in-depth interviews, but its findings should be interpreted with caution, understanding that the reported frequency of themes should be viewed in a comparative manner rather than taken at face value~\cite{fossey2002understanding}. This would  avoid potential misinterpretation or overgeneralization of the results.

Thirdly, we need to acknowledge the limitations associated with the sample size and demographics of our user study. The study was conducted with a specific group of participants, and, therefore, the findings may not fully represent the practices and perspectives of all AI practitioners. Our sample predominantly consisted of male participants, which aligns with the gender distribution reported in Stack Overflow's 2022 Developer Survey, where 92.85\% of professional developer respondents identified as male~\cite{stackoverflow2022survey}. Additionally, our participants were drawn from a large research-focused technology company. While the results may offer insights into practices within certain companies, they also serve as a case study for future research. 

Lastly, our qualitative results suggest indicators of ease of use for AI practitioners but does not provide direct information on the actual effectiveness of the guidelines. Understanding the impact of guidelines (or other AI toolkits~\cite{wong2023seeing}) requires long-term studies that consider multiple projects, with some utilizing the toolkit and others not. One potential avenue is to conduct observational studies with users of an AI system in a ``naturalistic setting''. Another approach is to use proxies such as measuring users' attitudes, beliefs, and mindset regarding ethical values before and after utilizing the guidelines. 
\section{Conclusion}
\label{sec:conclusion}

We proposed a method for generating a list of responsible AI guidelines that are grounded in regulations and are usable by different roles. The resulting 22 guidelines were integrated into an interactive tool and evaluated through a user study with 14 AI researchers, engineers, designers, and managers from a large technology company. Our participants found the guidelines well-aligned with their roles, enabling them to communicate complex ethical concepts in a structured manner. The guidelines are also grounded in ISOs and the EU AI Act articles, receiving positive feedback for being comprehensive. The usefulness of examples in guidelines was particularly noted as they enabled participants to reflect on their choices concerning ethical issues. As these guidelines are likely to become part of future responsible AI toolkits, it is important to implement features that provide users with time and space for reflection. Additionally, these toolkits should take users' reflections and roles into account to offer actionable recommendations tailored to a specific project, using, for example, large language models.
% \input{sections/4_DesignCards}

\begin{acks}

\end{acks}

% \nocite{*}
\bibliographystyle{ACM-Reference-Format}
\bibliography{main}

% \appendix
% \newpage
\appendix
\clearpage

\section{Additional Materials For the User Study}
\label{app:demographics-survey}
\begin{itemize}
    \item How old are you?
    \item What is your gender? [Male, Female, Non-binary, Prefer not to say, Open-ended option]
    \item How many years of experience do you have in AI systems?
    \item What's your educational background?
    \item In which country do you currently reside?
    \item What is domain or sector of your work? (e.g., health, energy, education, finance, technology, food)
    \item What is your current role?
    \item What kinds of AI systems do you work on? (e.g., machine learning, computer vision, NLP, game theory, robotics)
\end{itemize}

\begin{table}[h]
\centering

\caption{Constructed themes for the user study based on how our participants saw the application of guidelines, what worked well and what could have been improved.}
\label{tab:codebook-interviews}
% \resizebox{\textwidth}{!}{%
\begin{tabular}{ll}
\toprule
\textbf{Theme} & \textbf{Participants} \\ \midrule
\quad Raising awareness, facilitating self-learning & 12 \\
\quad Aligning with roles & 10 \\
\quad Aligning with regulations & 10 \\
\quad Providing helpful examples & 7 \\
\quad Engaging team members and external experts & 5 \\
\quad Maintaining the visual simplicity of the guidelines & 3 \\
\quad Documenting guidelines in a concise summary PDF & 3 \\
\quad Providing a systematic flow of information and guidelines & 2 \\
\end{tabular}%
% }
\end{table}

\clearpage

\section{Mapping Guidelines with EU AI Act Articles}
\label{app:mapping_guidelines}

\noindent\textbf{Article 6 (\emph{Classification rules for high-risk AI systems}):} It states that an AI system shall be considered high-risk when \emph{``it [the AI system] is intended to be used as a safety component of a product, or is itself a product''.} This article aligns with \textbf{guideline \#1} as it mandates the identification of an AI system's intended use to determine whether its use poses a low or high risk.
\smallskip

\noindent\textbf{Article 9 (\emph{Risk management system}):} 
It states that \emph{``a risk management system shall be established, implemented, documented and maintained throughout the entire lifecycle of a high-risk AI system''.} This article aligns with \textbf{guidelines \#1, \#3-5, and \#13} as it is about the identification of harms and risks of the AI system's intended use.
\smallskip

\noindent\textbf{Article 10 (\emph{Data and data governance}):} 
It states that \emph{``training, validation and testing data sets shall be subject to appropriate data governance and management practices''.} This article aligns with \textbf{guidelines \#8 and \#15-18} as it discusses the management and quality of data for training, validation, and testing, including aspects of diversity and minimizing biases.
\smallskip

\noindent\textbf{Article 11 (\emph{Technical documentation}):}  
It states that the technical documentation of a high-risk AI system shall \emph{``be drawn up before that system is placed on the market or put into service and shall be kept up-to date''}, and \emph{``provide national competent authorities and notified bodies with all the necessary information to assess the compliance of the AI system''.} This article aligns with \textbf{guidelines \#2, \#6, \#14} as it about documentation of the system and its contractual requirements, which may also be needed for obtaining ethical approvals.
\smallskip

\noindent\textbf{Article 12 (\emph{Record-keeping}):} 
It states that high-risk AI systems shall include \emph{``logging capabilities to enable the monitoring of the operation of the high-risk AI system with respect to the occurrence of situations that may result in the AI system presenting a risk''.} This article aligns with \textbf{guidelines \#6, \#9, \#10, and \#14} as it is about providing mechanisms for interpretable outputs and auditing, and improving the security of the system. 
\smallskip

\noindent\textbf{Article 13 (\emph{Transparency and provision of information to users}):} 
It states that \emph{``high-risk AI systems shall be designed and developed in such a way to ensure that their operation is sufficiently transparent to enable users to interpret the system's output and use it appropriately''.} This article aligns with \textbf{guidelines \#8-10, \#16-18, and \#20} as it is about quality, representativeness, and fit of training and testing datasets with the intended use.
\smallskip

\noindent\textbf{Article 14 (\emph{Human oversight}):} 
It states that \emph{``high-risk AI systems shall be designed and developed in such a way, including with appropriate human-machine interface tools, that they can be effectively overseen by natural persons during the period in which the AI system is in use''.} This article aligns with \textbf{guidelines \#9 and \#20} as it about ensuring human control over the system.
\smallskip

\noindent\textbf{Article 15 (\emph{Accuracy, robustness and cybersecurity}):} 
It states that \emph{``high-risk AI systems shall be designed and developed in such a way that they achieve, in the light of their intended purpose, an appropriate level of accuracy, robustness and cybersecurity, and perform consistently in those respects throughout their lifecycle''.} This article aligns with \textbf{guideline \#10} as it is about documenting the security of all system components.
\smallskip

\noindent\textbf{Article 16 (\emph{Obligations of providers of high-risk AI systems}):} 
It states that \emph{``providers of high-risk AI systems shall draw-up the technical documentation of the high-risk AI system''.} This article aligns with \textbf{guideline \#6} as it is about system documentation.
\smallskip

\noindent\textbf{Article 17 (\emph{Quality management system}):} 
It states that ``an AI system shall be documented in a systematic and orderly manner in the form of written policies, procedures and instructions''. This article aligns with \textbf{guidelines \#6, \#7, \#10, and \#14-18} because it is about documentation of all system components, including AI models and testing and validation procedures.
\smallskip

\noindent\textbf{Article 18 (\emph{Obligation to draw up technical documentation}):}    
It states that \emph{``providers of high-risk AI systems shall draw up the technical documentation ''.} This article aligns with \textbf{guideline \#6} as it is about system documentation.
\smallskip

\noindent\textbf{Article 20 (\emph{Automatically generated logs}):} 
It states that \emph{``providers of high-risk AI systems shall keep the logs automatically generated by their high-risk AI systems, to the extent such logs are under their control by virtue of a contractual arrangement with the user or otherwise by law''.} This article aligns with \textbf{guideline \#19} as it is about monitoring of the system.
\smallskip

\noindent\textbf{Article 29 (\emph{Obligations of users of high-risk AI systems}):} 
It states that users shall \emph{``monitor the operation of the high-risk AI system on the basis of the instructions of use.''}, and \emph{``inform the provider or distributor when they have identified any serious incident or any malfunctioning and interrupt the use of the AI system''.}  This article aligns with \textbf{guideline \#19} as it about monitoring of the system and utilizing guardrails or rollbacks.
\smallskip

\noindent\textbf{Article 50 (\emph{Document retention}):} 
It states that \emph{``the provider shall, for a period ending 10 years after the AI system has been placed on the market or put into service, keep at the disposal of the national competent authorities the technical documentation''.} This article aligns with \textbf{guideline \#6} as it about system documentation.
\smallskip

\noindent\textbf{Article 60 (\emph{EU database for stand-alone high-risk AI systems}):} 
It states that information contained in the EU database shall \emph{``be accessible to the public''} and \emph{``include the names and contact details of natural persons who are responsible for registering the system and have the legal authority to represent the provider''.} This article aligns with \textbf{guideline \#4} as it is about providing mechanisms for reporting system harms.
\smallskip

\noindent\textbf{Article 61 (\emph{Post-market monitoring by providers and post-market monitoring plan for high-risk AI systems}):} 
It states that \emph{``the post-market monitoring system shall actively and systematically collect, document and analyse relevant data provided by users or collected through other sources on the performance of high-risk AI systems throughout their lifetime''.} This article aligns with \textbf{guidelines \#12, \#14, \#15, \#19} as it is about data handling and model updates when the AI system is in use.
\smallskip

\noindent\textbf{Article 62 (\emph{Reporting of serious incidents and of malfunctioning}):} 
It states that \emph{``providers of high-risk AI systems placed on the Union market shall report any serious incident or any malfunctioning of those systems which constitutes a breach of obligations under Union law intended to protect fundamental rights to the market surveillance authorities of the Member States where that incident or breach occurred''.} This article aligns with \textbf{guideline \#4} as it is about incentivizing the reporting of system harms.
\smallskip

\noindent\textbf{Article 63 (\emph{Market surveillance and control of AI systems in the Union market}):} 
It states that \emph{``the national supervisory authority shall report to the Commission on a regular basis the outcomes of relevant market surveillance activities. ''.} This article aligns with \textbf{guideline \#4} as it about incentivizing the reporting of system harms.
\smallskip

\noindent\textbf{Article 64 (\emph{Access to data and documentation}):} 
It states that \emph{``access to data and documentation in the context of their activities, the market surveillance authorities shall be granted full access to the training, validation and testing datasets used by the provider, including through application programming interfaces (`API') or other appropriate technical means and tools enabling remote access''.} This article aligns with \textbf{guidelines \#16 and \#17} as it is about data documentation.
\smallskip

\noindent\textbf{Article 65 (\emph{Procedure for dealing with AI systems presenting a risk at national level}):} 
It states that \emph{``AI systems presenting a risk shall be understood as a product presenting a risk defined in Article 3, point 19 of Regulation (EU) 2019/1020 insofar as risks to the health or safety or to the protection of fundamental rights of persons are concerned''.} This article aligns with \textbf{guideline \#3} as it is about harms and risks identification.
\smallskip

\noindent\textbf{Article 67 (\emph{Compliant AI systems which present a risk}):} 
It states that if the AI system is compliant with the EU AI Act but still presents a risk to the health or safety of persons, the market surveillance authority \emph{``shall require the relevant operator to take all appropriate measures to ensure that the AI system concerned, when placed on the market or put into service, no longer presents that risk, to withdraw the AI system from the market or to recall it within a reasonable period, commensurate with the nature of the risk, as it may prescribe''.} This article aligns with \textbf{guideline \#5} as it is about mitigation strategies about the identified harms and risks.
\smallskip

\noindent\textbf{Article 69 (\emph{Codes of conduct}):} 
It states that \emph{``the Commission and the Board shall encourage and facilitate the drawing up of codes of conduct intended to foster the voluntary application to AI systems of requirements related for example to environmental sustainability, accessibility for persons with a disability, stakeholders participation in the design and development of the AI systems and diversity of development teams on the basis of clear objectives and key performance indicators to measure the achievement of those objectives''.} This article aligns with \textbf{guidelines \#2, \#11, \#21, \#22} as it is about the environmental assessment of the system, the ethical approvals obtained from ethics committees and boards, and the characteristics of the development team.

\end{document}